 \documentclass{emulateapj}  
 
\usepackage{graphicx}
\usepackage[dvips]{color}

\def\be{\begin{equation}}
\def\ee{\end{equation}}
\def\ba{\begin{eqnarray}}
\def\ea{\end{eqnarray}}

\def\12{{1\over 2}}

\def\msun{M_\odot}

\def\etal{{\it et~al.~}}

\def\ltsima{$\; \buildrel < \over \sim \;$}
\def\simlt{\lower.5ex\hbox{\ltsima}}
\def\gtsima{$\; \buildrel > \over \sim \;$}
\def\simgt{\lower.5ex\hbox{\gtsima}}

\shorttitle{Signatures of particle decay}
\shortauthors{Vasiliev \& Shchekinov}

\begin{document}

\title{Signatures of particle decay in 21 cm absorptions from first minihalos}

%% Use \author, \affil, and the \and command to format
%% author and affiliation information.
%% Note that \email has replaced the old \authoremail command
%% from AASTeX v4.0. You can use \email to mark an email address
%% anywhere in the paper, not just in the front matter.
%% As in the title, use \\ to force line breaks.

\author{Evgenii O. Vasiliev and Yuri A. Shchekinov}
\affil{Southern Federal University, Stachki Ave. 194, Rostov-on-Don, 344090 Russia}
\email{eugstar@mail.ru, yus@sfedu.ru}

%% Notice that each of these authors has alternate affiliations, which
%% are identified by the \altaffilmark after each name.  Specify alternate
%% affiliation information with \altaffiltext, with one command per each
%% affiliation.

%%%%%%%%%%%%%

%% Mark off your abstract in the ``abstract'' environment. In the manuscript
%% style, abstract will output a Received/Accepted line after the
%% title and affiliation information. No date will appear since the author
%% does not have this information. The dates will be filled in by the
%% editorial office after submission.

\begin{abstract}
The {imprint} of decaying dark matter (DM) particles in characteristics of the ``21 cm forest'' 
-- absorptions in 21 cm  from minihalos in spectra of distant radio-loud sources -- is considered  
{within a 1D self-consistent hydrodynamic description
of minihalos} from their turnaround point to virialization. {The most pronounced influence of decaying DM on evolution 
of minihalos is found in} the mass range $M=10^5-10^6\msun$, for which {unstable DM} with the current upper limit of the ionization rate 
$\xi_{L} = 0.59\times 10^{-25}$~s$^{-1}$ depresses 21 cm optical depth by an order of 
magnitude compared to the standard recombination scenario. {Even rather a} modest ionization $\xi \sim 0.3\xi_L$ practically 
``erases'' absorption features and {results in a considerable decrease} (by factor of more than 2.5) of the number of 
strong ($W_\nu^{obs} \simgt 0.3$~kHz at $z\simeq 10$) absorptions. {At such circumstances broad-band observations are to 
be more suitable for inferring physical conditions of the absorbing gas}. 
X-ray photons from stellar activity of initial episodes of 
star formation can compete the contribution from decaying DM only at $z<10$. Therefore, {when observed} 
{the 21 cm signal will allow to follow evolution} of decaying DM particles  
in the redshift range $z=10-15$. {Contrary, in case of non-detection of the 21 cm signal in the frequency range 
$\nu<140$ MHz a lower limit on the ionization rate from decaying dark matter can be established. }
\end{abstract}


%----------------------- Section 1 -------------------------------
\section{Introduction}

{ The standard 21-cm tomography of the Universe through observations of emission or absorption from the 
neutral intergalactic medium (IGM) against the cosmic microwave background (CMB) \citep{shaver99,sethiHI}, and
statistical studies of angular distribution of the 21 cm intensity \citep{tozzi,iliev02}, {are known to} suffer  
limitations from yet insufficient angular resolution of the existing and future radiotelescopes: e.g., {only} 
few arcseconds in the frequency range of interest ($\sim 150$ MHz) on LOFAR\footnote{http://www.lofar.org/index.htm},
which allows to distinguish only about hundreds of comoving kpc. Therefore only huge structures, such as
for example, HII regions formed by stellar clusters and quasars, and the large scale domain structure seem 
possible to be resolved.  On the contrary, measurements of absorptions in the 21 cm line to distant quasars -- the ``21 cm forest''
\citep{carilli02,furla02} -- is thought to provide information on small scale structures, such as 
individual dark matter minihalos, low mass galaxies and even stellar HII regions in the early Universe 
\citep{kumar,nathHI,furla02,furla06,ciardi21,mack11}.} {If so the 21 cm forest would serve as} a promising tool for studying the very beginning 
of the reionization epoch, when HII regions from different star forming minihalos have not yet {had} overlapped. 

{  Detailed study of the line profiles in the 21 cm forest has therefore attracted 
recently much attention. {In particular, dynamical effects on to the line profile from baryon
accretion \citep{ferrara11}, and from overall contraction during the formation of the minihalos } 
\citep{meiksin11,vs12} have been recently accounted, contrary to a fully static model with fixed dark matter and
baryonic profiles described initially in \citep{furla02,furla06}. Moreover, the 21 cm forest signal seems to carry
information on spatial distribution of dark matter in minihalos, and thus can shed light on ongoing discussion on
whether dark matter profiles in minihalos are cuspy as shown first in \citep{nfw}, or not as inferred by
\citep{burkert} from observations of local dwarf galaxies. The discussion has been recently exacerbated, on one side, by finding 
a steepening of the inner-slope of dark matter profiles in the higher mass ellipticals in clusters
\citep{sommer10,gnedin11}, and on the other, a flattening of profiles in the first protogalaxies,
\citep{mashch,tonini,dor06}. Detailed consideration of characteristics of the 21 
cm forest to firmly probe dark matter profiles at early epochs is therefore needed.}  

{ Considerable contamination can however come from stellar UV photons \citep{ciardi21,ferrara11} that 
ionize and heat gas in minihalos. Similar effects are expected from ionizing photons produced by decaying DM particles
\citep{sciama82,scott,sethiddm,kamiondecays,sneutrinos2,kasuya04a,kasuya04b,pierpaoli,belikov}, annihilation 
of dark matter \citep{chuzhoy07,nusser07,leptonddm,silk21ddm} and ultra-high energy cosmic rays {originating from 
decaying superheavy ($M_X \simgt 10^{12}$~GeV) DM particles \citep{berez,birkel,kuzmin,DN,dnnn}. 
The background radiation from these sources can change thermal and ionization
evolution of the intergalactic medium \citep{dodelson,biermannH2ddm,sh04,vas06,mapelli}, and affect through it
the 21 cm global signal and fluctuations \citep{furl06,sh07,chuzhoy07,nusser07,leptonddm,silk21ddm,natarajan}. 
More recently these sources have been strongly constrained 
\citep{kamion07,anihreion,delopeddm,kamion10,zhangddm,galli11anih,wmapanih}, though not fully excluded. 
In this paper we focus therefore on whether the imprints from them can be recognized in the 21 cm forest 
absorptions}.}

We assume a $\Lambda$CDM cosmology with the parameters $(\Omega_0,\Omega_{\Lambda},\Omega_m,\Omega_b,h)=(1.0,0.76,0.24,0.041,0.73)$ \citep{wmap}, for minihalos we assume a flat dark matter profile.

%----------------------- Section 2 ----------------------------

\section{Model Description}
\subsection{Evolution}

{ Dark matter profiles are described as suggested} by \citet{ripa}: the 
dark matter of $M_{DM} = \Omega_{DM} M_{halo}/\Omega_M$ is a truncated isothermal sphere with  
the truncation radius $R_{tr}$ evolving as in \citet{t97}, and a flat core of radius $R_{core}$; the ratio  
$\eta = R_{core} /R_{vir}$ is assumed 0.1 through simulations, as commonly  
used to mimic the evolution of a simple top-hat fluctuation \citep[e.g.][]{padma}.
Dynamics of baryons is described by a 1D Lagrangian scheme similar to that 
proposed by \citet{1d}; a reasonable convergence is found at a resolution of 
1000 zones over the computational domain. 

Chemical and ionization composition includes a standard set of species: 
H, H$^+$, H$^-$, He, He$^+$, He$^{++}$, H$_2$, H$_2^+$, D, D$^+$, D$^-$, HD, HD$^+$ and $e$, 
with the corresponding reaction rates from \citep{galli98,stancil98}. 
Energy equation includes radiative losses typical in primordial plasma:
Compton cooling, recombination and bremsstrahlung radiation, collisional 
excitation of HI \citep{cen92}, H$_2$ \citep{galli98} and HD \citep{flower00,lipovka05}. 

Calculations start at redshift $z = 100$. The initial parameters: 
gas temperature, chemical composition and other quantities -- are taken from 
simple one-zone calculations begun at $z = 1000$ with typical values taken at the end of 
recombination: $T_{gas} = T_{CMB}$, $x[{\rm H}] = 0.9328, x[{\rm H^+}] = 0.0672, 
x[{\rm D}] = 2.3\times 10^{-5}, x[{\rm D^+}] = 1.68\times 10^{-6}$ \citep[see references 
and details in][Table 2]{ripa}. 

%----------------------- Section 2 ----------------------------
\subsection{Ionization and heating}

Decaying DM contributes into ionization and heating of baryons \citep[see for recent discussion in] []{bertone}. 
The nature of decaying dark matter particles as well as the decay products are still under debates. 
Several candidates have been proposed: superheavy particles \citep{partrev}, axions \citep{axions1,axions2}, sterile
neutrinos \citep{sneutrinos1,sneutrinos2}, weakly interacting massive particles \citep[WIMPs, e.g.][]{jungman}.
The expected mass regime may range from very light of few keV (as sterile neutrino) and relatively light of few GeV \citep[e.g. ][]{kamiondecays}, 
to very heavy -- several TeV, as recently discussed by \citet{fr12}. Depending on the nature of the decaying particle, the
decay products may include $\gamma$-photons, electrons, positrons and other more exotic particles. For example, detection of positrons 
of tens of GeV by the PAMELA \citep{pamela} can be explained by the injection from annihilation and/or decay of dark 
matter \citep{pamelaanih,pameladecay}. It is clear however that through 
formation of particle showers the decays end when weakly interacting 
stable particles escape, while the other couple to baryons via electromagnetic interactions, ionize them and deposit  
energy.

The most generic form for the ionization rate from decaying DM particle is proposed by  
\citep{kamiondecays}
\be
I_e(z) = \chi_i f_x \Gamma_X { m_p c^2 \over h \nu_c}
\label{ix}
\ee
where $\chi_i$ is the energy fraction deposited into ionization \citep{shull}, $m_p$ is the 
proton mass, $f_x =\Omega_X(z)/\Omega_b(z)$, $\Omega_b(z)$, the baryon density parameter, 
$\Omega_X(z)$, the fractional abundance of decaying particles, $\Gamma_X$ is the decay rate, 
$h\nu_c$, the energy of Ly-c photons. { It is obvious that the contribution from decaying DM 
is determined by the product $\xi_i = \chi_i f_x \Gamma_X$ which masks the nature of dark matter particles}.

The corresponding heating rate can be written in the form \citep{kamiondecays}

\be
K = \xi_h m_p c^2
\label{kx}
\ee
where $\xi_h=\chi_hf_x\Gamma_X$, $\chi_h$ is the energy fraction depositing into {heating. By order of 
magnitude $\chi_i\sim\chi_h\sim 1/3$ for the conditions we are interested in \citep{shull}. }

Using the CMB datasets \citet{kamion07} have constrained the ionization rate associated with radiatively decaying 
dark matter as $\xi  \simlt 1.7\times 10^{-25}$~s$^{-1}$. An extended analysis of the data 
of Type Ia supernova, Ly$\alpha$ forest, large scale structure and weak lensing observations have lead 
\citet{delopeddm} to a stronger constraint: $\xi \simlt\xi_L = 0.59\times 10^{-25}$~s$^{-1}$.
It is worth noting that all datasets favor long-living decaying dark matter particles with the lifetime 
$\Gamma_X^{-1} \simgt 100$~Gyr \citep{delopeddm}. Further improvement is expected from the Planck satellite.
In this paper we consider models with the ionization rate within this limitation $\xi \le \xi_{L}$. 
{ As will be seen below such a constraint does not affect understanding of what would happen outside the 
limit $\xi>\xi_L$.} 

Equations (\ref{ix}) and (\ref{kx}) are written in assumption of a uniform production of 
ionizing photons by decaying DM with the density $\Omega_X$ without accounting its {  concentration in 
minihalos}. Virialized minihalos are 
transparent for sufficiently energetic photons for which $\tau(E) 
= r_{\rm vir} n_{\rm vir}  \sigma(E)<1$, where $r_{\rm vir}$ and $n_{\rm vir}$ are the radius and the mean density 
of a virialized halo, $\sigma(E)$ is the photoionization cross-section. The corresponding lower energy of escaping
photons is thus 

\be
E_l \simgt 300 {\rm eV}  \left({M\over 10^8 \msun }\right)^{1/9} 
             \left({1+z\over 10}\right)^{2/3}. 
\ee
{ 
In this sense our estimates correspond to a lower limit of ionization and heating rates.
}

The upper energy limit for the photons able to heat and ionize the IGM is determined 
such that they are absorbed in the IGM within the Hubble time, and can be readily estimated as $E_u\simlt 30$~keV
\citep[for a detailed description of the ``transparency window'' see in][]{kamiondecays}. In the energy range
$E_l<E<E_u$ photons are supposed to fill the IGM homogeneously except relatively small circumgalactic regions. 
{Note in this connection that even though each minihalo is a source of ionizing photons, their influence on the  
circumhalo baryons is obviously} negligible because of a very low optical depth $\tau(E)$ at the 
energies of interest.

{ 
Contrary to decaying DM stellar and quasi-stellar sources of ionizing radiation emerge only at redshifts $z<20$.  
In addition, they heat nearby surrounding gas, which then emits in 21 cm with a 
patchy, spot-like distribution on the sky, and thus the imprints from nonstellar and stellar sources can be 
discriminated (see discussion below in Sect. 4).}

%----------------------- Section 3 ----------------------------
\subsection{21 cm: optical depth and equivalent width}

The spin temperature of the HI 21 cm line is determined by atomic collisions and scattering 
of ultraviolet (UV) photons \citep{field,wout}. In our {calculations we used collisional coefficients from 
\citet{kuhlen} and \citet{liszt}. In general, decaying DM deposit into the budget of Ly$\alpha$ photons 
due to recombinations. However, in our present calculations we neglect UV Ly$\alpha$-pumping 
because it is small compare to collisional processes for suitable parameters of the decaying DM 
and the redshifts of interest \citep[see discussion in][]{sh07}. Moreover, the }
contribution from stellar Ly$\alpha$ background is small due to a 
sparsity of first stellar objects at redshifts $z=10-15$.

The optical depth along a line of sight at the frequency $\nu$ is 

\ba
\label{optd} 
 \tau_\nu = {3h_p c^3 A_{10}\over 32 \pi k \nu_0^2} \int_{-\infty}^{\infty}
   dx {n_{HI}(r) \over \sqrt{\pi} b^2(r) T_s(r)} \\ 
   \nonumber
   {\rm exp}\left[{-{[v(\nu) - v_l(r)]^2\over b^2(r)}}\right]
\ea
where $r^2 = (\alpha r_{vir})^2 + x^2$, $\alpha=r_\perp/r_{vir}$ is the dimensionless impact parameter, $v(\nu)=c(\nu-\nu_0)/\nu_0$, $v_l(r)$ 
is the infall velocity projected on the line of sight, $b^2 = 2kT_k(r)/m_p$ is the Doppler parameter.

The observed line equivalent width is determined as $W_\nu^{obs} =W_\nu / (1+z)$, where the intrinsic equivalent 
width is 

\be
 {W_\nu \over 2} = \int_{\nu_0}^{\infty}{(1-e^{-\tau_\nu})d\nu} 
                          - \int_{\nu_0}^{\infty}{(1-e^{-\tau_{IGM}})d\nu}
\label{eweq}
\ee
where $\tau_{IGM}$ is the optical depth of the background neutral IGM. 

%----------------------- Section 4 ----------------------------
\section{Results}

\noindent

%%%%%%%%%%%%%%%%%%%%%%%%%%%%%%%%%%%%%%%%%%%%%%%%%%%%%%
\begin{figure*}
\includegraphics[width=120mm]{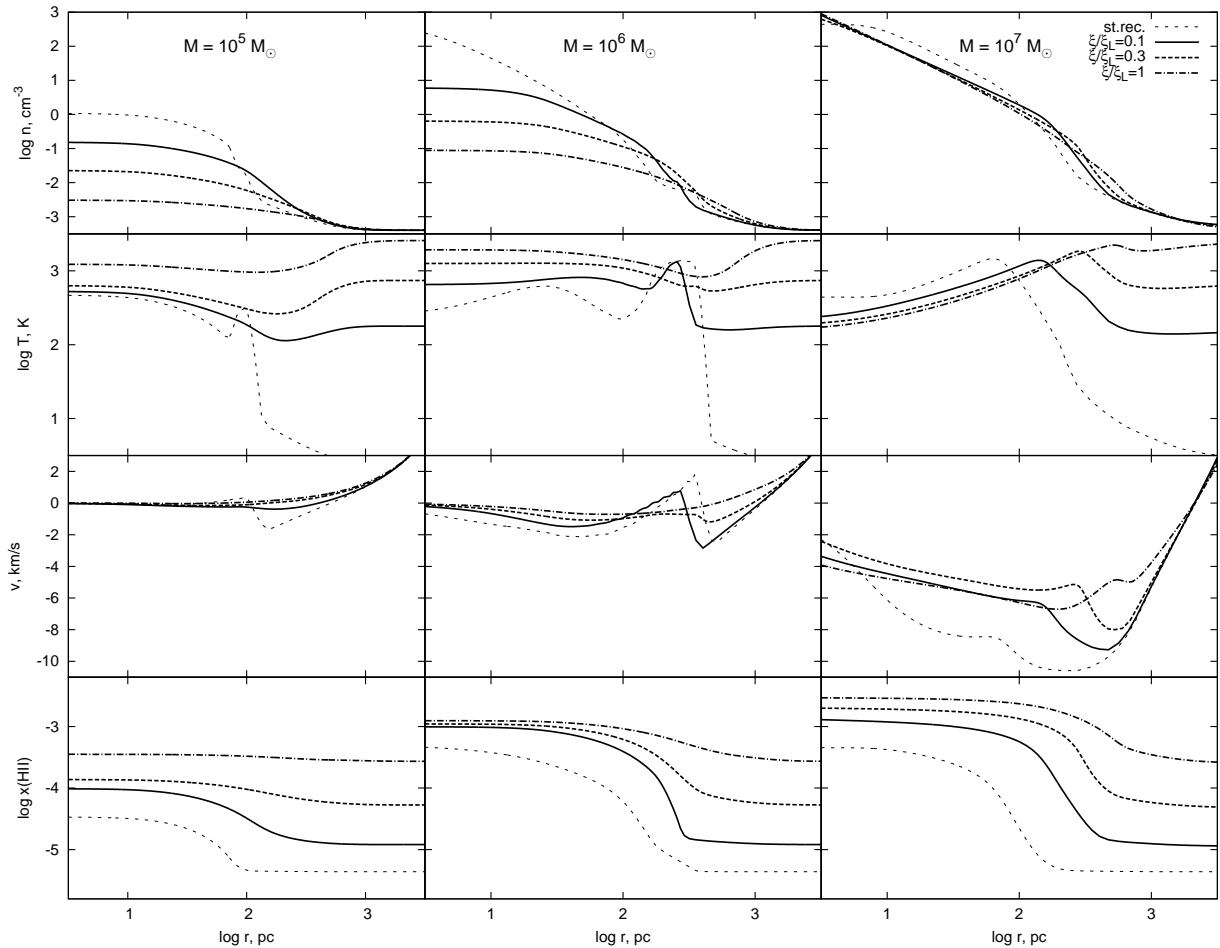}
\caption{
{
The radial density (upper), temperature (upper middle), velocity (lower middle) and 
ionized hydrogen fraction (lower panels) profiles
of halos $M=10^5\msun, \ 10^6\msun, \ 10^7\msun$ (from left to right panels, correspondingly) 
virialized at $z_{vir} = 10$  in the standard recombination model (dots)
and in the presence of decaying dark matter with $\xi/\xi_{L} = 0.1, 0.3, 1$ (solid, dash and
dot-dashed lines, correspondingly). 
}
}
\label{fig1}
\end{figure*}
%%%%%%%%%%%%%%%%%%%%%%%%%%%%%%%%%%%%%%%%%%%%%%%%%%%%%%

%\red{HERE}

\subsection{Dynamics of minihalos}

{We consider here only evolving minihalos with masses in the range $M=10^5-10^7\msun$ virialized 
at $z_v=10$.} 
At lower redshifts reionization by stellar and quasi-stellar sources comes in to play and contaminate the effects 
from decaying DM, while on the other hand, at higher redshifts the number of bright background radio sources 
decreases. {The choice of the halo mass range is motivated by the fact that halos of $M=10^5\msun$ lie below the limit where 
the baryons can efficiently cool and form stars. Only inside minihalos with $M \simgt 2\times 10^6\msun$ star formation can potentially 
occur in the sense that they collapse at $z_{vir}=10$, i.e. the 
gas density in the most inner shell reaches the critical value $10^8$~cm$^{-3}$ when 3-body reactions turn on to form H$_2$ \citep{ripa}}. 
On the other side, minihalos of $M=10^7\msun$ do eventually collapse and form stars and thus represent the opposite 
dynamical regime. {The results we present here illustrate therefore the typical features of 21 cm absorptions 
produced by non-starforming ($M=10^5\msun$), marginally starforming ($M=10^6\msun$) 
and starforming ($M=10^7\msun$) minihalos. }  

Heating from decaying dark matter (hereafter, DDM heating) weakens the accretion rate of baryons on to 
minihalos. Therefore, in comparison with the evolution of minihalos in the standard scenario one can 
expect smaller baryon mass and higher temperature within the virial radius of a minihalo. The influence 
of decaying particles depends on the minihalo mass and the rate of the decay energy deposited in gas, $\xi$. 

Figure~\ref{fig1} show radial profiles of density (upper), temperature { (upper middle)}, velocity { (lower 
middle) and ionized hydrogen fraction} (lower panels) in minihalos $M=10^5, 10^6, 10^7\msun$ virialized at 
$z_{vir} = 10$, (from left to right panels, correspondingly); 21 cm absorption profiles from collapsing minihalos 
the standard recombination model can be found in \citep{vs12}. It is clearly seen that the IGM temperature under 
DDM heating can approach or even exceed the virial temperature of minihalos: DDM with $\xi \simgt 0.1 \xi_{L}$ heats 
the IGM up to $T\simeq 250$ K which is only half of the virial temperature of a $M=10^5\msun$ halo, while $\xi \simgt
0.3\xi_{L}$ elevates the IGM temperature close to the virial temperature of $M=10^6\msun$ minihalos. It results in
weakening of baryon accretion. 
Thus the most obvious and important difference {between evolving minihalos in the standard recombination regime 
and the presence of decaying dark matter} is the absence of an accretion shock wave in minihalos with $M\simlt
10^6\msun$, as obviously seen on the lower panels of Figure~\ref{fig1}. 

{The radial density, temperature and velocity profiles in small mass ($M\simlt 10^6\msun$) minihalos in the presence of DDM heating 
look similar. The profiles\footnote{As for minihalos with $M=10^7\msun$ the collapse occurs earlier than formal virialization, we present 
the results for not later than the collapse redshift, i.e. for $z=11$.} in massive halos ($M=10^7\msun$) remain close to those in the 
standard recombination scenario (see dash lines in Figure~\ref{fig1}) inspite of a strong 
DDM heating.} Thus contrary 
to the halos of smaller mass where gravitation is too weak to overcome additional heating, massive halos,  
$M\geq 10^7\msun$, instead remain able to support the accretion rate on a practically unchanged level. It is obvious that 
within the limit $\xi\leq \xi_L$ minihalos with $M=10^7\msun$ represent the least massive halos sensitive to an additional 
from DDM. However, as readily seen, further increase of $\xi$ results in a practically proportional growth of the IGM temperature 
$T_{_{IGM}}\propto\xi$, and consequently suppresses the accretion rate approximately as $T_{_{IGM}}^{-3/2}$, which in turn 
increases the lower limit of halo mass insensitive to DDM heating. 

{  Ionization from the decayng DM is considerable: it is seen from Fig. 1 that in the low-mass range  
($M\leq 10^6\msun$) the fractional ionization within minihalos and outside them is nearly invariant in models 
with $\xi\simgt 0.3 \xi_L$.} The ionization rate of $\sim 10\xi_L$ can affect the evolution of 
minihalos with masses of classical dwarf galaxies $M\sim 10^9\msun$.

%%%%%%%%%%%%%%%%%%%%%%%%%%%%%%%%%%%%%%%%%%%%%%%%%%%%%%%%%%%
\begin{deluxetable*}{ccccc}
\tablewidth{0pt}
\tablecaption{Optical depth in the center of line at virialization}
\tablehead{
\colhead{M, $\msun$} & 
\colhead{$\xi/\xi_L = 0$} & 
\colhead{$\xi/\xi_L = 0.1$} & 
\colhead{$\xi/\xi_L = 0.1, I_e = 0$} & 
\colhead{$\xi/\xi_L = 0.1, K = 0$} }
\startdata
$10^5$ &  0.15 & 0.074 & 0.09 & 0.12\\
%\hline
$10^6$ &  0.37 & 0.074 & 0.083 & 0.33 \\
%\hline
$10^7$ &  0.068 & 0.1 & 0.081 & 0.078
\enddata
\end{deluxetable*}
%%%%%%%%%%%%%%%%%%%%%%%%%%%%%%%%%%%%%%%%%%%%%%%%%%%%%%%%%%%

\subsection{Optical depth}

%%%%%%%%%%%%%%%%%%%%%%%%%%%%%%%%%%%%%%%%%%%%%%%%%%%%%%
\begin{figure*}
\includegraphics[width=110mm]{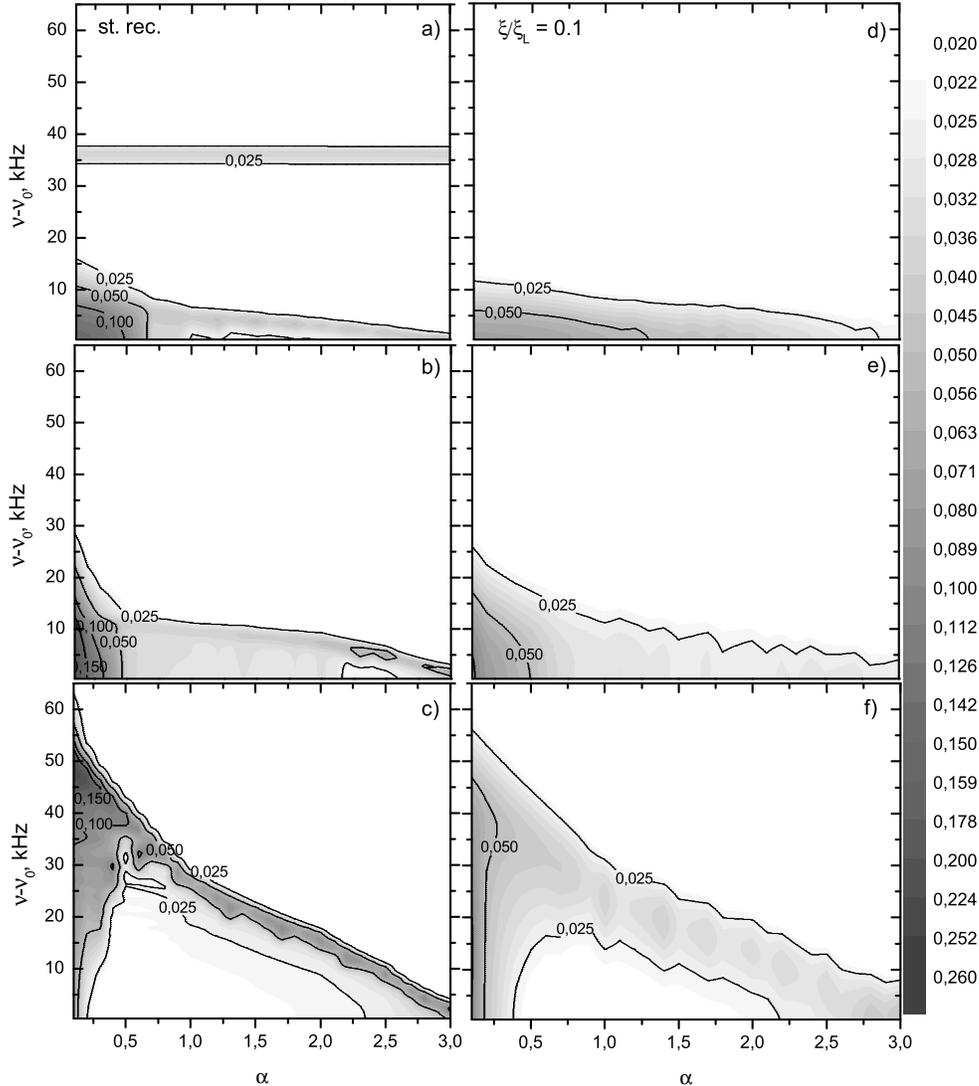}
\caption{
The dependence of the {  21cm} optical depth on frequency and impact parameter $\alpha=r_\perp/r_{vir}$ for halos 
$M=10^5\msun, \ 10^6\msun, \ 10^7\msun$ (from top to bottom, correspondingly) virialized at $z_{vir} = 10$,
%\footnote{For $M=10^7\msun$ the lowest redshift is $z\simeq11$, see Section~5.1 for details.}
in the standard recombination model (left column of panels) and in the presence of decaying dark matter with 
$\xi/\xi_{L} = 0.1$ (right column).  
}
\label{fig2}
\end{figure*}
%%%%%%%%%%%%%%%%%%%%%%%%%%%%%%%%%%%%%%%%%%%%%%%%%%%%%% 

New evolutionary features of minihalos introduced by decaying dark matter are to be manifested in optical depths and
equivalent widths of the 21 cm line. Figure~\ref{fig2} shows the maps of optical depth on the ``impact parameter --
frequency'' plane in the standard recombination model (left column of panels) and in the presence of decaying DM with 
$\xi/\xi_{L} = 0.1$. The analysis of optical depth in the 21 cm line from minihalos in the standard  recombination
scenario can be found in \citep{ferrara11,meiksin11,vs12}, so here we will speak only about the models with decaying 
DM. In addition, we will discuss only models with a single value of $\xi$ because the other are either similar or have 
no remarkable features and can be readily scaled. 

An obvious result is that decaying DM heavily suppresses optical depth. In the low-mass end 
$M=10^5-10^6\msun$ the center of line optical depth ($\nu-\nu_0\simlt 10$~kHz) for small impact factors 
($\alpha\simlt 0.5$) at $\xi = 0.1\xi_L$ is less than half as compared to the standard case. A small peak in 
optical depth at $\nu-\nu_0\simeq 40$~kHz for $M=10^5~\msun$ disappears \citep[the origin of this peak is discussed 
in ][]{vs12}. { At a higher impact factor $\alpha \simgt 1$ the optical depth remains insensitive to $\xi \simlt 
0.1\xi_L$.} However, higher $\xi$ can suppress optical depth down to the background value 
$\tau \sim 10^{-3}$ for all $\alpha$: for $M=10^5~\msun$ this occurs when $\xi \simgt 0.3\xi_L$, and
for $M=10^6~\msun$ it can be met for $\xi \simgt \xi_L$. Thus, the decaying DM with $\xi \simgt 0.3\xi_L$ 
practically erases 21 cm absorption features from small mass minihalos.

Massive minihalos, $M=10^7\msun$, demonstrate more complex frequency-dependence of the optical depth: it does 
not peak in the center of line, the maximum is instead at $\nu-\nu_0 \sim 45-50$~kHz. Such a horn-like dependence
originates from a strong accretion in massive halos \citep{ferrara11,vs12}. In the presence of decaying dark matter 
the range of impact factors with a horn-like profile heavily shrinks even for $\xi \sim 0.1\xi_L$. The centre of
line optical depth ($\nu-\nu_0\simlt 30$~kHz) at small $\alpha\simlt 0.1$ becomes flat and insensitive to $\xi$: it
stays at the level $\sim 0.05$ even for $\sim \xi_L$. At higher $\alpha$ the optical depth merges to the
background value.

{  It is clear that both the heating and the ionization from the decaying DM contribute to the decrease of optical 
depth. In order to evaluate their relative contributions we performed calculations of two sets of models: in the first 
set we turned off the ionization from the decaying DM $I_e=0$ leaving the heating unchanged, while in the second set 
the ionization remains unchanged, while heating is turned off $K=0$. The results are shown in Table 1, where the first
column shows the center-of-line optical depths $\tau_{0}$  
in the standard ionization scenario, in the second column $\tau_{0}$ is shown for the model with the decayng DM
particles treated self-consistently, the second and the third columns correspond to the models with $I_e=0$ and $K$
unchanged, and $I_e$ unchanged and $K=0$, respectively. An obvious trend of increased $\tau_{0}$  
in the shortened models with either $I_e=0$ or $K=0$ is caused by an inhibition of the total
contribution from the decaying DM. It is also clearly seen that the influence of the ionization $I_e$ is mainly not 
due to a decrease of the fraction of  HI in virialized halos which is negligibly small: $\Delta x({\rm HI})=-x({\rm
HII})\sim -10^{-3}$. It stems predominantly from an enhanced Compton heating beginning from the turnaround point.
Therefore both the ionization and the heating decrease the optical depth via the additional heating. Separate
contributions from them depend on the minihalo mass: for instance, for $10^5~\msun$ halos they comprise approximately 
1/3 for the ionization and 2/3 for the heating, although their contributions are not independent and the combined
influence is not simply a sum of the two separate: the additional heating from these sources change the temperature,
density and velocity profiles which are involved in the optical depth (\ref{optd}) nonlinearly.        
}

\subsection{Equivalent width}

%%%%%%%%%%%%%%%%%%%%%%%%%%%%%%%%%%%%%%%%%%%%%%%%%%%%%%
\begin{figure*}
\includegraphics[width=120mm]{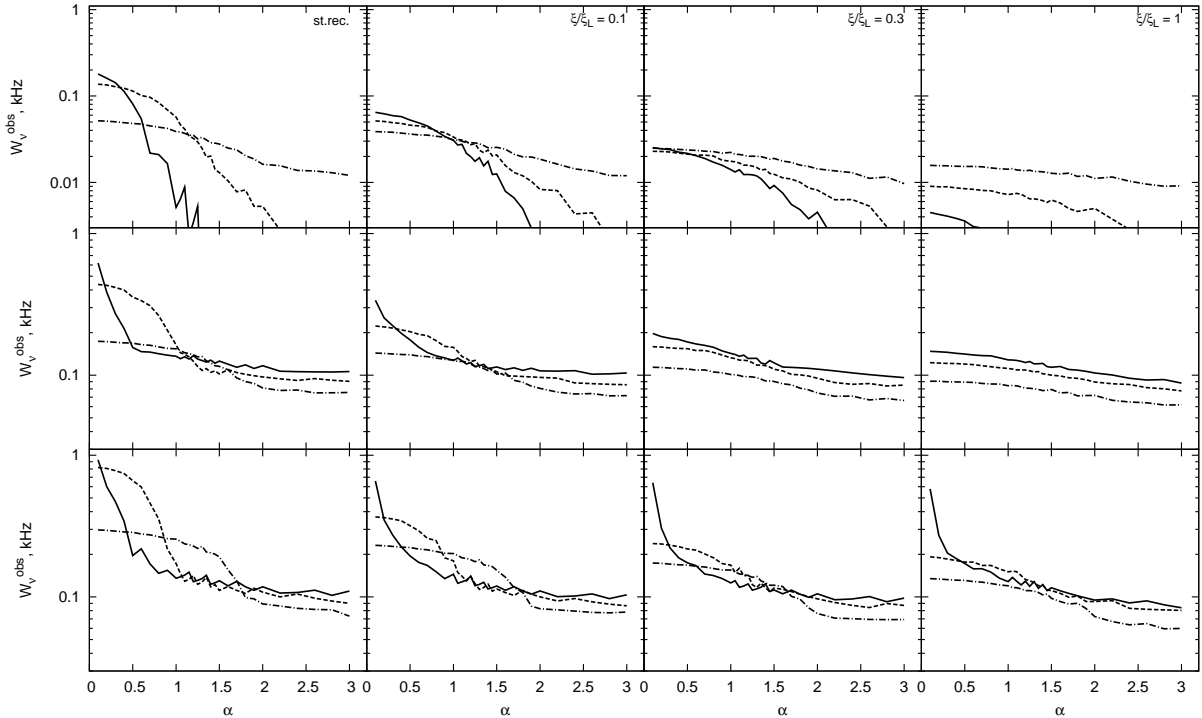}
\caption{
The equivalent widths versus the impact parameter $\alpha$ in the observer's restframe for minihalos with masses
$M=10^5,\  10^6,\  10^7\msun$ (from top to bottom) at redshifts $z = 15.5, 12$ and $z=z_{vir}=10$ (dot-dash, dash 
and solid lines, correspondingly) in the standard recombination model and in the presence of decaying dark matter 
with $\xi/\xi_{L} = 0.1, 0.3, 1$ (from left to right, correspondingly). Note that the $y$-axes are different between 
the panels.
}
\label{fig3}
\end{figure*}
%%%%%%%%%%%%%%%%%%%%%%%%%%%%%%%%%%%%%%%%%%%%%%%%%%%%%%

Figure~\ref{fig3} demonstrates evolution of the radial profiles of the equivalent widths measured at the observer's
restframe. As expected it decreases significantly with $\xi$ except those for the halos of $M=10^7\msun$: for 
instance, low-mass halos $M=10^5\msun$ at $z=10$ (solid lines) show decrease of $W_\nu^{\rm obs}$ from $\sim 0.2$~kHz
for the standard recombination to a negligible value $\simlt 0.02$~kHz for $\xi \geq 0.3\xi_{L}$ within
$r_\perp/r_{vir}\simlt 0.5$. At $z = 15.5, 12$ (dash and dot-dash lines) $W_\nu^{\rm obs}$ hardly can be resolved on 
future telescopes and may only deposit into the total signal 
in broad-band observations \citep{ferrara11}.

More massive halos -- $M\simgt 10^6\msun$, show near the virialization ($z=10$) strong equivalent widths 
$W_\nu^{\rm obs} \simgt 0.2$~kHz even at a relatively high ionization $\xi \simlt 0.3\xi_{L}$ within the whole 
halo $\alpha\simlt 1$, at $\alpha\simlt 0.5$  $W_\nu^{\rm obs}$ can reach even $\sim 0.5-0.9$~kHz. In the higher redshift range,
$z=15.5, 12$, $M\simgt 10^6\msun$ can produce such strong only in the standard recombination scenario. More massive halos,
$M=10^7\msun$, form strong absorptions ($W_\nu^{\rm obs}\simgt 0.5$~kHz) even for high ionization rate $\xi \sim \xi_L$,
though only in a narrow range of impact factor $\alpha \simlt 0.1$. {As a consequence, the decaying DM leads to a considerable
decrease of the number of strong absorption lines: for $\xi \simgt 0.3\xi_{L}$ sufficiently strong lines form only in
high-mass ($M\geq 10^7\msun$) halos nearly along the diameter with a small probability to contribute into absorptions. 
Similar effect stems from the action of the stellar X-ray background at lower redshifts \citep{furla02,ferrara11}, see also Sec. 4. }

\begin{figure}
\includegraphics[width=80mm]{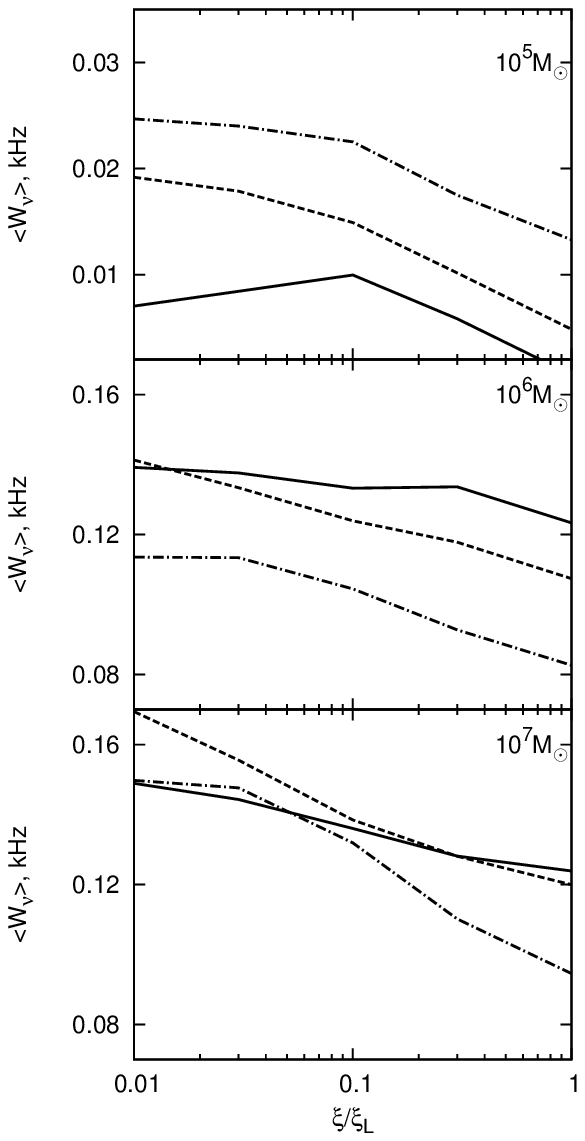}
\caption{
The dependence of the radially averaged equivalent width for minihalos with mass 
$M=10^5,\  10^6\msun,\  10^7\msun$ (from top to bottom panels, correspondingly) 
at redshifts $z = 15.5, 12$ and 10 (dot-dash, dash and solid 
lines, correspondingly) on the ionization rate $\xi/\xi_L$; in the logarithmic 
$x$-axis $\xi/\xi_L = 0.01$ is set to correspond to $\xi/\xi_L = 0$, i.e. the standard recombination.
Note differences between the panels on $y$-axes scales.}
\label{fig8}
\end{figure}

The sensitivity of future telescopes is at least an order of magnitude lower than the flux limit needed to separate 
spectral lines from individual halos, i.e. to observe with spectral resolution of $\Delta\nu\sim 1$~kHz. Therefore,
\citet{ferrara11} have proposed broadband observations with lower resolution. In this case the measured quantity is the average signal
from different halos lying on a line of sight. Accordingly, we introduce a radially averaged equivalent width
\be
{ 
\langle W_\nu^{obs}\rangle = {2\over 9} \int\limits_0^{3r_{vir}}{W_\nu^{obs}(r_\perp)r_\perp dr_\perp \over r_{vir}^2}.
}
\ee
Figure~\ref{fig8} presents the dependence
of $\langle W_\nu^{obs}\rangle$ on $\xi/\xi_L$. The increase of $\langle W_\nu^{obs}\rangle$ for $M=10^5\msun$ at $z=10$
and $\xi/\xi_L \leq 0.1$ is explained by a widening of $W_\nu^{obs}$ { in the peripheric region $r_\perp/r_{vir}\simlt 2$
(upper panels in Figure~\ref{fig3}). For higher masses and redshifts the radially averaged $W_\nu^{obs}$ decreases with
$\xi/\xi_L$ by factor of $\simlt 2$ and apparently remains sufficient to distinguish effects from the ionization
caused by decaying DM particles.  }

%%%%%%%%%%%%%%%%%%%%%%%%%%%%%%%%%%%%%%%%%%%%%%%%%%%%%%
%%%%%%%%%%%%%%%%%%%%%%%%%%%%%%%%%%%%%%%%%%%%%%%%%%%%%%

In previous studies theoretical spectrum of the 21 cm forest has been simulated within the assumption of 
steady-state minihalos with fixed profiles corresponding to the virialization \citep{furla02,ferrara11}. 
When dynamics is taken into account, minihalos lying on a given line of sight are, in general, at different 
evolutionary stages, and have different density, velocity and temperature profiles. Moreover, Press-Schechter 
formalism cannot anymore be used for description of their mass function as soon as they are lying far from the
virialization. It makes problematic to describe correctly the number density of minihalos at each evolutionary 
stage. 
{  An additional complication stems from continuous structure formation. While major mergers involving merging  
minihalos with $M\simgt 0.5M_h$ occur on times as long as
twice of the Hubble time at $z=10$, less destructive mergers with $M\simlt 0.25 M_h$ proceed within 
$\simlt 0.7$ of the Hubble time \citep{lacey}. This time is shorter than free-fall time in halos with 
$M\sim 2\times 10^6~\msun$, thus making estimates based on the assumption of the virial equilibrium  
unapplicable to less massive halos. 
}
Possible formation of stars in massive evolved minihalos also introduces additional complication. Overall, the 
analysis of theoretical spectrum within a statistical simulation becomes exceedingly cumbersome. More relevant 
picture can be obtained from high-resolution cosmological gas dynamic simulation, although current resolution 
with $\Delta M\sim 10^6~M_\odot$ seems not to be sufficient.

However, with the upcoming low frequency interferometers (LOFAR and SKA) such modeling is apparently excessive. 
Indeed, the absorption line profiles and the spectral features from separate minihalos located at $z=10$ are 
of 1 to 5 kHz in width -- close to the {spectral} resolution limit 1 kHz. As mentioned in \citep{ferrara11} the corresponding sensitivity requires enormously bright background sources. {In order to resolve spectral 
lines of $\Delta\nu\sim 1$~kHz with a typical for the SKA telescope ratio of the aperture-to-system 
temperature $A_{eff}/T_{sys}=5000$~m$^2$~K$^{-1}$ and a signal-to-noise ratio $S/N=5$ the integration 
time $\sim 30$~days and the minimum flux density about 500~$\mu$Jy are needed \citep{ferrara11}. This is 4 times lower than 
the flux density for an object similar to bright radio-loud galaxy Cygnus~A \citep{carilli02}, but an order 
of magnitude higher than typical GRB afterglow flux \citep{ferrara11}. Note, in addition, that there are no confirmed 
QSOs at $z>7.5$, and moreover the mass of a black hole in the Cygnus A galaxy is about $2.5\times 10^9\msun$ \citep{cygabh}, 
which should apparently be a very rare object at $z\simgt 10$. GRB afterglows seem to be more promising background sources 
for the 21 cm absorption study. In this case the integration time is restricted by the duration of a bright phase of afterglow, 
$\sim 100$~days \citep{frail03}. Thus, low resolution or broadband observations can be a better alternative. 

In broadband observations suppression of optical depth 
in 21 cm caused by decaying particles manifests as a factor of 2--4 decrease of absorption in the frequency 
range $\nu<140$ MHz where contribution from low mass minihalos at $z\geq 10$ dominates. Indeed, the number 
of halos in the low-mass range $10^5-10^7\msun$ at $z=10$ scales as $n(M,z)\sim M^{-1}$. The probability of 
minihalos to intersect a line of sight is proportional to $n(M,z)\times(\alpha r_{vir}(M))^2$, resulting in 
a decrease of the number of strong absorption lines with $W_\nu^{\rm obs} \simgt 0.3$~kHz by factor of at least 
2.5 for $\xi/\xi_{L} = 0.3$ and more than 4.5 for $\xi/\xi_{L} = 1$. Such a decrease inevitably cancels the 
average 21 cm forest signal from low mass halos. { It results in turn in 
a several times decrease of the mean flux decrement $D_A$, and correspondingly the detectability limit 
in broadband observations. Strong suppression of the optical depth at $\xi/\xi_{L} \simgt 0.3$
%(if we would like to achieve flux density limit $\sim 70$~$\mu$Jy with $A_{eff}/T_{sys}=5000$~m$^2$~K$^{-1}$ 
%and frequency bandwidth $\Delta\nu=5$~MHz) 
would require time integration scales longer than the duration of a bright phase 
of GRB afterglows.} Low-mass ($M\sim 10^5~\msun$) minihalos produce weak absorptions, $W_\nu^{\rm obs} 
\simlt 0.2$~kHz even for the standard ionization scenario. 

Thus, as far as the characteristic scales corresponding to such broadband observations are of the order 
of hundreds of kpc, an immediate consequence is that the advantages expected initially from the 21 cm forest
observations become not that obvious. However, once observations are made to cover a wide range of frequencies
corresponding to redshifts 10 to 15, they would in principle provide information about how the influence from decaying 
dark matter varies with redshift. {More explicitly, a non-detection of absorptions from 21 cm in the 
frequency range $\nu<140$ MHz can in principle serve to bound the ionization rate from decaying DM \it{from below}: 
$\xi/\xi_{L} \simgt 0.3$, unless contribution from stellar sources becomes competitive to ionize low mass halos.}

%----------------------- Section 5 --------------------------
\section{Relative contribution from particle decays and stellar X-ray background}

{  As mentioned above the contribution of X-ray photons from decaying particles into ionization and heating 
of the IGM can compete the one from X-ray produced be early population of binaries and/or black holes. Let us 
estimate here comparatively these two contributions. For this purpose we assume the X-ray ($>0.2$~keV) 
luminosity of high-$z$ galaxies to be comparable to that inferred for nearby starburst galaxies \citep{gilfanov04} 
\be
 L_X = 3.4\times 10^{40} f_X \left( { SFR \over 1~\msun~ {\rm yr^{-1}} } \right) ~ {\rm erg~s^{-1}}
\label{lumx}
\ee
where SFR is the star formation rate, and $f_X$ is a correction factor accounting for the unknown 
properties of X-ray emitting sources in the early universe. { Rather a weak constraint on $f_X$ inferred from the 
WMAP CMB optical depth $\simlt 10^3$ \citep{pritchard2008} is much higher than 
found  in the nearby universe, $f_X \simeq 0.3-1$ \citep{gilfanov04,mineo12}. From the signal 
in stacked Chandra images at the position of $z\sim 6$ galaxies \citep{treister11}, the soft
X-ray background is argued to limit the $f_X$ factor by $\simgt 2-5$ at $z\ge 1-2$,
though allows $f_X$ to increase sharply up to $\sim 100$ at $z>5$ \citep{dijkstra12}. However, further rigorous 
analysis did not confirm such a high $f_X$ \citep{willott11,fiore12}, and show that it remains invariant at least
within $z=0-4$ \citep{cowie12}. Thus, no reason are seen to assume $f_X$ much higher than that in the local universe.}

Assuming that the star formation rate is proportional to the rate at which matter collapses into galaxies
\citep{furla06}, the total normalized X-ray emissivity $\epsilon_X$ can be written as 
\be
{2 \epsilon_X \over 3 k n H(z) } = 5\times 10^4 f_X \left[ {f_*\over 0.1} {df_{coll}/dz\over 0.01} 
                                                            {1+z\over 10} \right], 
\label{epx}
\ee
here $df_{coll}/dz$ is the fraction of baryons collapsed to form a protogalaxy per unit redshift, $f_\ast$, the
fraction of baryons converted into stars in a single star formation event, normalization numbers are given in
\citep{furla06}. The corresponding heating rate can be found as $K_\ast=\chi_h\epsilon_X$, which can be now compared 
to the heating rate associated with decaying particles 
\be
{2K\over 3 k H(z) } = 8.9\times 10^4 (\xi/\xi_L) \left({1+z\over 10}\right)^{-2/3}. 
\label{kxx}
\ee
The condition that these two sources heat the IGM equally is therefore 
\be
0.54 f_X\chi_h \left[ {f_*\over 0.1} {df_{coll}/dz\over 0.01} \left({1+z\over 10}\right)^{5/3} \right] = (\xi/\xi_L). 
\ee
Thus, in order that the heating from stellar X-ray background were equal to the heating from decaying particles 
one would require for the X-ray correction factor $f_X \simeq 5.4(\xi/\xi_L)$ for $1+z=10$ and typical parameters 
for the star formation. At $\xi/\xi_L>0.1$ such value seems to be exceedingly higher than follows from observations 
of starburst galaxies in the local Universe with $f_X\simeq 0.3$ \citep{gilfanov04}. Moreover, the linear approximation
(\ref{lumx}) is valid only for a sufficiently massive systems with the star formation rate $\simgt 10~\msun$ yr$^{-1}$,
while in the low end of SFR the correction factor can be one to two orders lower. At the redshifts of interests $z\sim
10$ only halos with masses $M_h\simlt 10^7~\msun$ can be star-forming \citep{loeb10}, for which the SFR is at most 
$\sim 10^{-3}~\msun$ yr$^{-1}$ for typical numbers characterizing the formation of stars in the early Universe:
$f_\ast=0.1$, $df_{coll}/dz\sim 0.01$. Such a low star formation rate corresponds however to more than factor of 10
weaker proportion between the star formation and the X-ray luminosity, i.e. $f_X\ll 0.1$ \citep[see Fig. 1 
in][]{gilfanov04}. One may conclude therefore that at redshifts $z\geq 10$ heating of atomic hydrogen in virializing
minihalos can stem only from the decaying particles, unless the star formation efficiency $f_\ast$ and the amount of 
a collapsed baryons $df_{coll}/dz$ grows towards higher $z$ unprecedentedly. Without overwhelming assumptions the
conclusion about a dominance of decaying particles in heating looks most plausible. 

{ As mentioned above higher $f_X$ can in principle be met at $z\sim 10$ \citep{dijkstra12}, in which case the 
contribution from decaying particles decreases at $z\simlt 10$, though at higher redshifts they still dominate 
due to an obvious sharp drop of star formation. The different redshift dependence of the stellar X-ray and the decaying
particle heating rates (\ref{kxx}), may be used to discriminate between them.
In general though, both higher $\xi$ and $f_X$ lead to smaller number of absorption lines. As a consequence 
the decrement $D_A$ decreases, and the integration time increases such that even GRB afterglows cannot serve to observe the 
21 cm forest. }
} 

It is obvious that strong stellar X-ray sources, such as, e.g., SN remnants or massive X-ray binaries, ionize 
and heat the IGM locally, even though the X-ray background from them can be weaker than from decaying particles. 
Indeed, the mean free path of photons with $E\sim 0.3$~keV at $z\sim 10$ is about 1~Mpc/$(1+\delta)$, where $\delta$ 
is the density perturbation. The heating rate at a distance $D$ from a source with X-ray luminosity $L$ [erg~s$^{-1}$]
is
\ba
\Gamma \sim 6\times 10^{-28} \left({L\over 10^{40}}\right) \left({10 ~{\rm kpc} \over D}\right)^2\chi_h  \times \\
\nonumber 
                            \times \left({E/E_H\over 20}\right)^{-3}\exp(-\tau) ~{\rm erg~s^{-1}},
\ea
and the corresponding ionization rate is
\ba
I \sim 2\times 10^{-17} \left({L\over 10^{40}}\right) \left({10 ~{\rm kpc} \over D}\right)^2 \times \\ 
\nonumber
                            \left({E/E_H\over 20}\right)^{-3}\chi_i\exp(-\tau) ~ {\rm s^{-1}}. 
\ea
It is readily seen that in the presence of heating from decaying particles the zone of influence of a star-forming dwarf galaxy lies within the radius 
\be
D\simlt 10 \left( { SFR \over 1~\msun~ {\rm yr^{-1}} } \right)^{-1/2}(\xi/\xi_L)^{-1/2}~{\rm kpc}.
\ee
As far as a typical distance between minihalos $d\sim 30(1+z/20)$~kpc, only in the beginning of reionization 
epoch dwarf galaxies can dominate heating and ionization of hydrogen in neighbor minihalos only when the ionization
from the background X-ray photons from decaying particles is relatively weak, $(\xi/\xi_L)<0.1$.

%----------------------- Section 6 --------------------------
\section{Conclusions}

In this paper we have considered the influence of decaying dark matter particles on the HI 21 cm absorption 
features from low mass minihalos: $M=10^5,\  10^6\msun,\  10^7\msun$, virialized at $z_{vir} = 10$. 
We used a 1D self-consistent hydrodynamic approach to study their evolution, and followed the absorption
characteristics from the turnaround to the virialization of minihalos. 
We have found that

\begin{itemize}
\item due to additional heating from decaying particles thermal and dynamical evolution of minihalos 
show pronounced differences from those in the model without particle 
decays (i.e. with the standard recombination scenario), {  and can be distinguished in the signal 
formed at redshifts $z>10$}; 

\item {  the additional heating and ionization} strongly suppress optical depth in the 21 cm line: they practically ``erase'' 
the 21 cm absorptions from minihalos with $M=10^5-10^6\msun$ even at a relatively modest ($\xi \simgt 0.3\xi_L$)
ionization rate; the horn-like dependence of the optical depth found for minihalos  
with $M \sim 10^6 - 10^7\msun$ in the standard recombination scenario \citep{ferrara11} almost disappears even 
at a lower ionization rate $\xi \simgt 0.1\xi_L$. {  In total, approximately $1/3$ of this suppress stems from the ionization, 
while the rest $\sim 2/3$ is due to the heating. The contribution from the ionization is caused by an enhanced Compton heating 
on additional electrons, rather then by decreased fraction of HI.}

\item the suppress of the magnitude of 21 cm absorption by the ionization from decaying DM  
particles, and the widening of the lines by dynamical effects -- baryon accretion and dark matter contraction --
necessitate broadband observations, in which case the advantages of the 21 cm forest to probe reionization 
on small scales become less obvious as compared to the traditional 21 cm global signal and 21 cm angular distribution;

\item stellar contribution at such redshifts, $z\geq 10$, is negligibly small, so that a weakening of 
the 21 cm forest if observed can be only due to decaying dark matter particles. 

\item the equivalent width of the 21 cm absorption line decreases significantly when $\xi$ increases, and 
the number of strong absorption lines $W_\nu^{obs} \simgt 0.3$~kHz at $z=10$ drops by more than 2.5 to 4.5 times
depending on $\xi$; such a decrease inevitably erases the averaged 21 cm forest signal from low mass halos 
($M \sim 10^5- 10^6\msun$, i.e. the frequency range $\nu<140$ MHz) in future broad-band observations. 
{Within a conservative assumption that $f_X$ does not grow enormously and remains at higher redshifs $z\simgt 10$ 
comparable to its value in the local Universe, such decrease of the EW can be attributed only to the decaying 
DM. As far as it occurs at $\xi>0.3\xi_L$ the deficiency of 21 cm forest absorptions at these frequencies can serve to 
put \it{a lower limit} on the ionization rate from unstable DM.}

\end{itemize}

\acknowledgments

The authors thank M. Gilfanov for valuable and helpful comments, the anonymous referee for critical remarks,
and Shiv Sethi for discussions.
This work is supported by the Federal Agency of Education (project codes RNP 2.1.1/11879).
EV acknowledges support from the "Dynasty" foundation, the RFBR through the grant 12-02-00365 and
the Russian federal task program "Research and operations on priority directions of development of 
the science and technology complex of Russia for 2009-2013" (state contract 14.A18.21.1304).  

%----------------------- Section L -------------------------------

%------------------------ figures --------------------------------

\end{document}